\documentclass[10pt,apl,twocolumn,floatfix]{revtex4}
\usepackage{amsmath}
\usepackage[dvips]{graphicx}

\begin{document}

\title{Quantum random number generator based on twin beams}
\author{Qiang Zhang$^{1}$, Xiaowei Deng$^{1}$, Caixing Tian$^{1}$, and
Xiaolong Su$^{1,2}$}
\email{suxl@sxu.edu.cn}
\affiliation{$^{1}$State Key Laboratory of Quantum Optics and Quantum Optics Devices, \\
Institute of Opto-Electronics, Shanxi University, Taiyuan, 030006, People's
Republic of China\\
$^{2}$Collaborative Innovation Center of Extreme Optics, Shanxi University,\\
Taiyuan, Shanxi 030006, People's Republic of China}

\begin{abstract}
We produce two
strings of quantum random numbers simultaneously from the intensity
fluctuations of the twin beams
generated by a nondegenerate optical parametric oscillator. Two strings of
quantum random numbers with bit rates up to 60 Mb/s are extracted simultaneously with
a suitable post-processing algorithm. By post-selecting the identical data from two raw sequences
and using a suitable hash function, we also extract two
strings of identical quantum random numbers. The obtained random
numbers pass all NIST randomness tests. The presented scheme shows the
feasibility of generating quantum random numbers from the
intensity of a macroscopic optical field.
\end{abstract}

\maketitle

Random numbers have significant applications in science and engineering \cite%
{QRNG2016,Ma2016}, such as cryptography, statistical analysis, numerical
simulation, etc. There are two main categories of random numbers, which are
pseudo random numbers and true random numbers. Pseudo random numbers are
generated with a given software algorithm and finite length seed, which is
easy to achieve a high bit rate, but powerless in some stringent occasions.
True random numbers, which are generated from the measurement of unpredictable
physical processes, are more secure and reliable. There are various methods to
produce true random numbers, such as chaotic systems \cite%
{Chan2012,Li2016,Butler2016}, thermal noise in electronic circuits \cite%
{Petrie2000}, and optical noise of superluminescent LEDs
\cite{Li2011}.

True randomness is an essential part of quantum mechanics. A quantum random
number generator (QRNG) exploits the inherent randomness of a quantum event
to produce true random numbers. Several optical QRNGs have been proposed and
demonstrated, such as QRNGs based on photon counting \cite%
{Rarity1994,Jen2000,Stef2000,Piro2010,Dynes2008}, attenuated pulse \cite%
{Wei2009,Bisadi2015,Sti2015}, phase noise of laser \cite%
{Guo2010,Qi2010,Zhu2011,Xu2012,Zhou2015,Nie2015}, quantum vacuum
fluctuations \cite{Shen2010,Gab2010,Symul2011,Zhu2012,Shi2016}, Raman
scattering \cite{Bustard2011,Bustard2013} and optical parametric oscillators
(OPO) \cite{Marandi2012,Okawachi2016}. Up to now, the QRNGs with bit rates up
to Gbit/s have been achieved \cite{Zhu2011,Xu2012,Nie2015,Symul2011}.
Recently, a QRNG based on the photonic integrated circuit has been
demonstrated \cite{Abellan}, which shows the feasibility of integrated QRNGs.

In the previous QRNG based on an OPO, the phase of the macroscopic field is
used to produce random numbers, where two independent cavities of the same
output power are used and two output fields interfered at a beam splitter
\cite{Marandi2012}. In another QRNG based on an OPO, the frequency-degenerate
bi-phase state of a dual-pumped degenerate OPO in a silicon nitride
microresonator is used to produce random numbers \cite{Okawachi2016}. The twin beams generated in the parametric down-conversion process are
well-known to have intensity correlation \cite%
{Heid1987,Sou1997,Wang1999,Gao1998,Lau2003,Zhang2003} and quantum
entanglement \cite{Villar2005,Su2006,Jing2003}. In the spontaneous
parametric down conversion, a nonlinear medium converts a photon at frequency
$\omega _{0}$ into two photons at frequency $\omega _{s}$ and $\omega _{i}$
with $\omega _{0}=\omega _{s}+\omega _{i}$. In a nondegenerate optical
parametric oscillator (NOPO), a type-II crystal is inserted into an optical
cavity. When a NOPO is operated above threshold, the vacuum fluctuations are amplified and the
continuous entangled twin beams are obtained. The vacuum fluctuations are a fundamental quantum effect, which cannot be influenced by a potential adversary. The previous QRNG based on an OPO only outputs one string of random numbers. However, based on the twin beams generated by NOPO, two strings of quantum random numbers can be produced simultaneously.

In this letter, we demonstrate an efficient method to produce
quantum random numbers from twin beams, which are generated by a
NOPO. The true randomness is guaranteed by the inherited quantum
fluctuations of the twin beams. The intensity fluctuations of the
twin beams are measured directly by two photodetectors in the time
domain respectively. With post-processing, we extract two strings of
quantum random numbers simultaneously. Based on the quantum
correlation of the twin beams, we also extract two strings of
identical quantum random numbers by post-selecting the identical
bits from two raw sequences. Using Toeplitz hashing, the
self-correlation of each individual random string is reduced. The
obtained random numbers pass all NIST randomness tests.

\begin{figure}[tbp]
\begin{center}
\includegraphics[width=80mm]{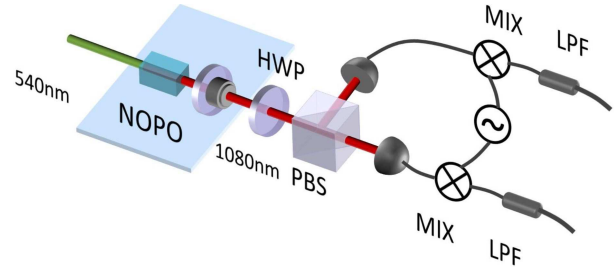}
\end{center}
\caption{Schematic of the QRNG based on twin beams. HWP: half-wave
plate, PBS: polarization beam-splitter, MIX: mixer, LPF: low-pass filter.}
\end{figure}

Figure 1 shows the schematic of the QRNG based on twin beams. A
continuous-wave laser beam at wavelength of 540 nm is used as the
pump beam of a NOPO. The NOPO consists of an $\alpha $-cut type-II
KTiPO4 (KTP) crystal and a concave mirror, which is a semimonolithic
configuration. The front face of the KTP is coated as the input
coupler and the concave mirror with 50 mm curvature serves as the
output coupler of the twin beams. The front face of the KTP crystal
is coated with a transmission of 7\% at 540 nm and high reflectivity
at 1080 nm. The output coupler is coated with a transmission of
12.5\% at 1080 nm and high reflectivity at 540 nm. The cavity length
is 54 mm. In our experiment, the threshold of the NOPO is about 50
mW.

The cavity length of the NOPO is locked on the pump resonance using a
feedback servo system. With a pump power of 80 mW, the NOPO emits two
continuous orthogonally polarized twin beams with near degenerate
wavelength at 1080 nm. The twin beams are separated by a
polarization beam splitter and then focused on a pair of detectors
with carefully balanced amplifications. A half-wave plate is
inserted before the polarization beam splitter. When the
polarization of the twin beams is rotated by an angle of 45 degrees,
the measured noise in the intensity difference is the shot noise
limit (SNL). When the polarization of the twin beams is rotated by
an angle of 0 degrees, the measured noise in the intensity
difference is the intensity difference spectrum of the twin beams \cite%
{Heid1987}.

Theoretically, the measured intensity difference
spectrum of the twin beams generated by the NOPO is expressed as \cite%
{Heid1987,Su2006}
\begin{equation}
S(\Omega )=S_{SNL}(1-\frac{\eta \xi }{1+\Omega ^{2}\tau _{c}^{2}})
\end{equation}%
where $S_{SNL}$ is the shot noise limit (usually normalized to 1), $\Omega $
is the measured noise frequency; $\tau _{c}$\ is the cavity storage time; $%
\eta $ is the total efficiency of the detection system (including
quantum efficiency of photodiode and transmission efficiency of twin
beams); $\xi =T/(T+\delta ) $ is the output coupling efficiency of
NOPO , in which $T$ is the transmission coefficient of the output
coupling mirror, and $\delta $ is the loss of the
cavity. With parameters $\Omega =4$ MHz, $\tau _{c} =0.0196$ $\mu $s, $%
\eta =89.3\%$, and $\xi =95.3\%$, the theoretical noise reduction for intensity difference of twin beams is 8.1 dB. Figure 2(a) shows the measured intensity difference noise of twin beams in the frequency domain. The intensity difference noise between the twin beams is 6.3 dB below the SNL around 4 MHz, where the electronic noise is about 27.6 dB lower than the SNL. The difference between the experimentally measured squeezing and the theoretical squeezing mainly comes from the imperfection of the experimental system, for example, the thermal effect of the NOPO and fluctuation of the locking system.

To exploit the correlated intensity fluctuations of the twin beams
to generate quantum random numbers, we use the detected intensity
noise at 4 MHz with a bandwidth of 600 kHz. The ac output of each
detector is mixed with a 4 MHz sinusoid signal and then
filtered by a low-pass filter whose cutoff frequency is 300 kHz \cite%
{Shen2010,Gab2010,Symul2011}. The intensity noises of twin beams are sampled
and digitized with a 8-bit ADC (National Instrument 5153) using the sampling frequency of 10 MHz. The bit rate of the produced quantum random numbers depends on the bandwidth of the NOPO, photodetector and low-pass filter used in the measurement device. If a broadband NOPO and photodetector are used, and the photocurrents are filtered with a broadband low-pass filter, the bit rate of the produced random numbers can be increased.

Figure 2(b) shows the measured intensity noises of twin beams in the time
domain. The intensity fluctuations of twin beams are correlated for about
75\%. The statistical histograms of the digitized intensity noises of twin
beams are shown in Fig. 2(c) and 2(d), respectively. It is obvious that
the distribution of the measured intensity noise of each one of the twin
beams is Gaussian.

\begin{figure}[tbp]
\begin{center}
\includegraphics[width=80mm]{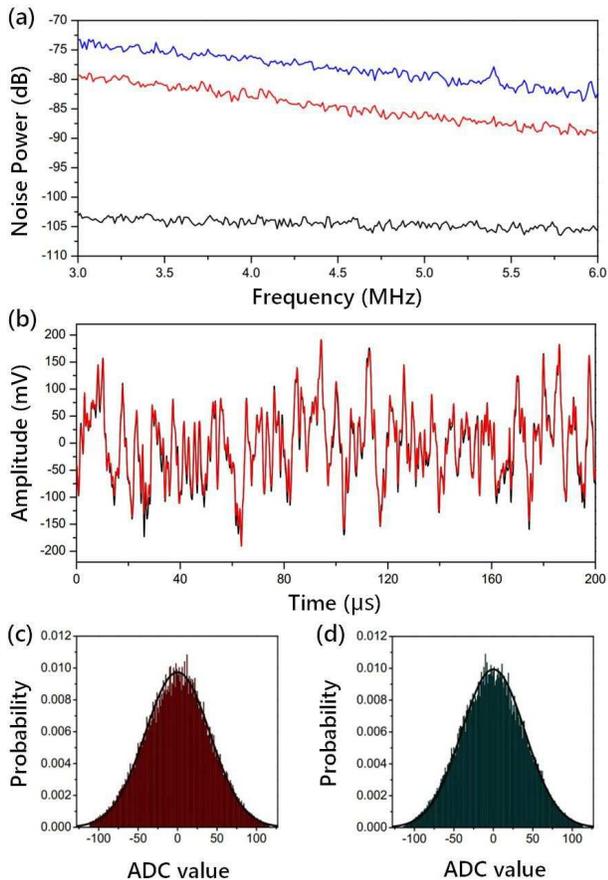}
\end{center}
\caption{(a)
Measured intensity difference noise in the frequency domain. The
traces from top to bottom are the SNL, intensity difference noise
and electronic noise, respectively. (b) Measured intensity noises in
the time domain. Red and black lines correspond to ac outputs of
the two detectors with a sample size of 2000 points, respectively. (c)
- (d) Statistical histograms corresponding to each one of the
measured intensity noises of the twin beams with 20,000 data
digitized by the 8-bit ADC, respectively. The mathematically fitting
curves are plotted with the Gaussian outlines.}
\end{figure}

\begin{figure}[tbp]
\begin{center}
\includegraphics[width=80mm]{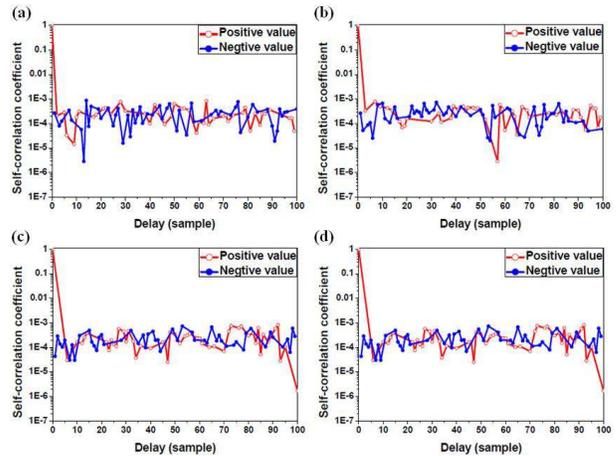}
\end{center}
\caption{Results
of self-correlation analysis. Data size is 80 Mbits. (a)-(b) Results corresponding to each string of quantum random numbers with
bit rate of 60 Mb/s, respectively. (c)-(d) Results
corresponding to each of two strings of identical quantum random
numbers, respectively.}
\end{figure}

In order to extract quantum random numbers, we apply the min-entropy
to quantify the quantum randomness of the obtained raw data. In some
sense, the min-entropy is a measure of the maximum amount of
information that can be obtained under a single attack
\cite{Wayne2010}. The min-entropy is exploited to quantify the
extraction ratio between the raw random bits and the final random
bits given a probability distribution of \{0, 1\}$^{N}$
\cite{Zhang2016}, which is evaluated as
\begin{equation}
H_{\min }(X)=-\log _{2}(\max_{X\in \{0,1\}^{N}}\Pr [X=x]).
\end{equation}%
For a given sequence $X$, the min-entropy is determined by the
sample point with maximal probability $P_{\max }=\max_{X\in
\{0,1\}^{N}}\Pr [X=x]$.

It is reasonable to assume the quantum noise of the intensity fluctuation is
perfectly random over all frequencies and independent of the classical
noise. We describe the variance of the ac output voltage from the detector
as $\sigma _{total}^{2}=\sigma _{quant}^{2}+\sigma _{c}^{2}$, where $\sigma
_{quant}^{2}$\ and $\sigma _{c}^{2}$\ are variances of the quantum signal
and the classical signal, respectively \cite{Symul2011,Shi2016}. We obtain $%
\sigma _{total}^{2}$\ as $4768.44$ mV$^{2}$ for the measured intensity
noise. In our system, the classical noise mainly comes from the electronic
noise of the detector, which is about $3.18$ mV$^{2}$. Therefore, the
quantum variance is $\sigma _{quant}^{2}=4765.26$ mV$^{2}$. Assuming the
quantum signal follows Gaussian distribution,\ the corresponding maximum
probability of the raw data is $P_{\max }=0.00993296$. Consequently, the
quantum min-entropy in our experiment is estimated to be $6.65$ bits\ per
sample \cite{Xu2012}.

The raw random data cannot pass any randomness tests, mainly because the
classical noise is mixed into the raw data and the sample points follow
the Gaussian distribution rather than the uniform distribution \cite{Xu2012}%
. In order to distill the randomness of the raw random data, we utilize the
Toeplitz hashing to eliminate the classical noise and improve the
statistical quality of the random numbers \cite{Xu2012,Zhang2016}. Given $%
m\times n$ binary Toeplitz matrix, $m$ random bits are extracted by
multiplying the Toeplitz matrix with $n$ raw bits. We choose $m=1024$ and $%
n=1360>1024\times 8/6.65=1232$ to obtain nearly perfect random bits. We use $%
n+m-1=2383$ pre-stored true random bits as seed to construct the
Toeplitz matrix. Therefore, two strings of quantum random numbers
are extracted simultaneously at rates up to $60$ Mb/s.

Based on the correlated intensity fluctuations of the twin beams, we
also distill two strings of identical quantum random numbers by
changing the post-processing algorithm. We post-select identical
bits between the raw data sequences and discard the different bits.
After post-selection, 70\% of the raw random bits are selected as
the input string of the Toeplitz hashing. We set $m=1024$ and
$n=1920$ to construct the Toeplitz matrix. Consequently, two strings
of identical quantum random numbers are extracted simultaneously at
rates up to $29.8$ Mb/s. The bit rate of identical quantum random
numbers depends on the quantum correlation between intensity
fluctuations of the twin beams. The higher the quantum correlation of the twin
beams, the higher the bit rate; this is because there are more
identical bits to be selected between two raw data sequences (less
different bits are discarded).

The self-correlation of the obtained quantum random numbers is verified by
self-correlation coefficient $R(k)$\ of a sequence $X$\thinspace\ which is
defined as \cite{Xu2012,Shi2016}
\begin{equation}
R(k)=\frac{E[(X_{i}-\mu )(X_{i+k}-\mu )]}{E[(X_{i}-\mu )^{2}]}.
\end{equation}%
where $E[\cdot ]\ $is the expected value operator, $k$\ is the sample delay
and $\mu $\ is the mean of $X$. Figure 3 shows that the self-correlation of
the random numbers is reduced by post-processing. The average values ($k\neq
0$) of Fig. 3(a)-3(d) are $-3.64\times 10^{-5}$, $%
-1.09\times 10^{-5}$, $3.59\times 10^{-6}$, $3.59\times 10^{-6}$,
respectively.

\begin{figure}[tbp]
\begin{center}
\includegraphics[width=80mm]{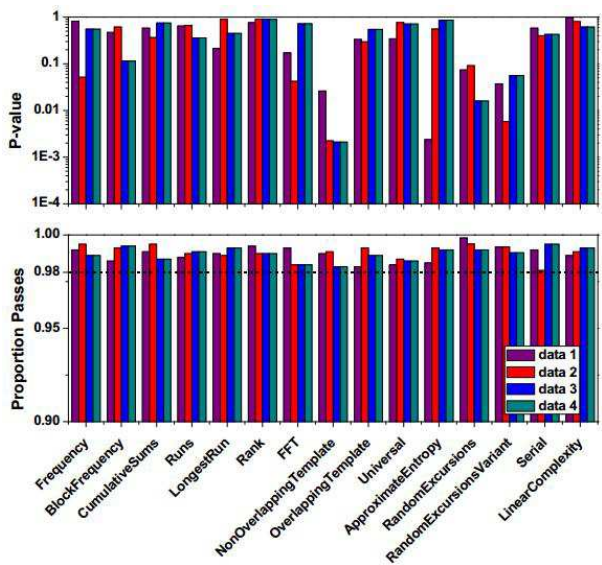}
\end{center}
\caption{Results of NIST statistical test suites. Using 1000 samples of
1Mb and significance level $\protect\alpha $=0.01. For \textquotedblleft
Pass\textquotedblright , the $P$-value (uniformity of $p$-values) should be
larger than 0.0001, and the proportion should be within the range of 0.99$\pm
$0.0094329. For the tests that produce multiple $P$-values and proportions,
the worst case is shown. Data 1 and data 2 show the results corresponding to
each string of quantum random numbers with a bit rate of 60 Mb/s,
respectively. Data 3 and data 4 show the results corresponding to each one of
two strings of identical quantum random numbers, respectively.}
\end{figure}

The NIST test \cite{Rukhin2010} is widely considered as one of the most
stringent randomness test suites. It has 15 statistical tests to evaluate
the performance of a given random number generator. Each test output a
$p$-value. A significance level $\alpha \ $is chosen for the test. For
cryptographic application $\alpha \ $is mostly set as $0.01$. A tested
sequence is considered to pass the test if the $p$-value $\geq \alpha $;
otherwise, the sequence appears to be nonrandom. We record two random
sequences of 1 Gbits. Each sequence is chopped into 1000 smaller sequences
for the NIST test. Each test calculates 1000 $p$-values, we use the chi-square
test to calculate the final $P$-value, which indicates the uniformity of
$p$-values. Figure 4 shows the results of NIST statistical test suites for two
strings of quantum random numbers and two strings of identical quantum
random numbers. All of them pass these tests.

In summary, we demonstrate an efficient method to produce quantum random
numbers from intensity fluctuations of twin beams. The true randomness is
guaranteed by the inherited quantum fluctuations of the twin beams. We observed 75\% quantum correlation  between the intensity fluctuations of the twin beams in the time domain. The intensity fluctuations of the twin beams are measured directly
by two photodetectors and two strings of quantum random numbers with bit
rates up to 60 Mb/s are extracted simultaneously with a suitable
post-processing algorithm. By post-selecting identical bits, we also extract
two strings of identical quantum random numbers with bit rates of 29.8 Mb/s
using the same device. The obtained random numbers pass all NIST randomness
tests, which confirms the randomness of the generated random numbers.

The advantage of the presented scheme is that two strings of random
numbers can be extracted simultaneously, especially two strings of
identical quantum random numbers which are extracted by post-selection.
The bit rate of the quantum random numbers produced in this scheme
is limited by the bandwidth of the preparation and measurement
systems of the twin beams. To obtain a higher bit rate of two strings
of identical quantum random numbers, twin beams with higher quantum
correlation are required, which means that lower intracavity losses
of NOPO and a detection system with better performance are required.
It has been shown that twin beams can be generated by an on-chip
monolithically integrated optical parametric oscillator \cite{Dutt},
which shows the possibility of an integrated QRNG based on twin
beams. Our work also shows the possibility to use entangled lights
to generate quantum random numbers.

This research was supported by National Natural Science Foundation of China (NSFC) (11522433, 61475092) and National Basic Research Program of China
(2016YFA0301402).

\end{document}